\documentstyle[sprocl,epsf]{article}

\def\Journal#1#2#3#4{{#1} {\bf #2}, #3 (#4)}

\def\PLB{{\em Phys. Lett.}  B}

\def\PRD{{\em Phys. Rev.} D}
\def\ZPC{{\em Z. Phys.} C}

\begin{document}

\title{\vskip-2.5cm{\baselineskip14pt
\centerline{\normalsize\hfill BUTP--99/06}
\centerline{\normalsize\hfill April 1999}
}
\vskip.7cm
Hadronic contribution to $\alpha(M_Z^2)$ and the anomalous magnetic
  moment of the muon\footnote{Talk
        presented at the {\it $17^{th}$ 
          International Workshop on Weak Interactions and Neutrinos (WIN 99)},
        Cape Town, South Africa, January 24--30, 1999.}}

\author{
Matthias Steinhauser}
\address{Institut f\"ur Theoretische Physik, Universit\"at
      Bern, CH-3012 Bern, Switzerland\\E-mail: ms@itp.unibe.ch}


\maketitle

\abstracts{
In this contribution recent developments are discussed which lead to a
significant reduction of the error for $\alpha(M_Z^2)$ and $(g-2)_\mu$.
}

The impressive amount of data mainly collected at LEP, SLC and TEVATRON has
resulted in a very high precision sometimes even below the per mille level.
The combination with accurate computations has made it possible to predict the
mass of the top quark, $M_t$, before its actual experimental discovery.
A similar scenario is nowadays applied to the Higgs boson mass. However, the
dependence of the radiative corrections on $M_H$ is much weaker than on $M_t$
which has the consequence that the limits are much more loose. Nevertheless
it is possible to derive an upper bound on $M_H$ of roughly 300~GeV
at 95\% confidence level~\cite{mhval}.

A crucial role in the indirect determination of $M_H$ is taken over by the
error on the electromagnetic coupling at the scale $M_Z$, $\alpha(M_Z^2)$.
Adopting the value given in Ref.~\cite{EidJeg95}
and expressing the error
in terms of uncertainties in the weak mixing angle it actually 
dominates in the comparison with the other error sources~\cite{GamRADCOR98}
(putting aside the error on $M_H$, of course).
The authors
of~\cite{EidJeg95} performed a very conservative analysis which exclusively
relies on
rather imprecise data below an center-of-mass energy, $\sqrt{s}$, of 40~GeV.
Only for $\sqrt{s}\ge40$~GeV perturbative QCD is used.

A similar situation can be found in the case of the anomalous magnetic moment
of the muon. The error cited in~\cite{EidJeg95} reads $\pm153\times10^{-11}$
which has to be compared with the aimed experimental error of
$\pm40\times10^{-11}$ in the E821 experiment at Brookhaven.

These numbers show that it is very important to improve both the error on
$\alpha(M_Z^2)$ and on $(g-2)_\mu$. In the recent months several different
options 
have been suggested which significantly reduced the uncertainties. In this
contribution they are briefly discussed and compared at the end.

The electromagnetic coupling at the scale $M_Z$ is given by
\begin{eqnarray}
\alpha(s) &=& \frac{\alpha(0)}
      {1-\Delta\alpha_{\rm lep}(s)
        -\Delta\alpha^{(5)}_{\rm had}(s)
        -\Delta\alpha_{\rm top}(s)}
\,.
\end{eqnarray}
with $\alpha=\alpha(0)=1/137.0359895$. The leptonic~\cite{Ste98} and the top
quark~\cite{KueSte98} contribution are both known up to the three-loop order
where the errors are negligible:
\begin{eqnarray}
\Delta\alpha_{\rm lep}(M_Z^2) \,\,=\,\, 314.98 \times 10^{-4}
\,,
&& 
\Delta\alpha_{\rm top}(M_Z^2) \,\,=\,\, -0.70 \pm 0.05 \times 10^{-4}
\,.
\end{eqnarray}
The evaluation of the hadronic contribution requires the computation of the
following dispersion integral.
\begin{eqnarray}
\Delta\alpha_{\rm had}^{(5)}(M_Z^2)
&=&
-\frac{\alpha M_Z^2}{3\pi}\,\mbox{Re}\,
\int_{4m_\pi^2}^\infty\,{\rm d}
s\,\frac{R(s)}{s\left(s-M_Z^2-i\epsilon\right)}
\,,
\label{eq:delaldisp}
\end{eqnarray}
with $R(s)=\sigma(e^+e^-\to\mbox{hadrons})/\sigma(e^+e^-\to\mu^+\mu^-)$.
It is not possible to use perturbation theory for $R(s)$ in the whole energy
region. Thus one has to rely --- to a more or less large extend --- on
experimental data.

Similarly the anomalous magnetic moment receives contributions from different
areas~\cite{CzaMar98}:
\begin{eqnarray}
\left(\frac{g-2}{2}\right)_\mu &\,\,\equiv\,\,& a_\mu \,\,=\,\,
a_\mu^{\rm QED} + a_\mu^{\rm had} + a_\mu^{\rm ew}
\,.
\end{eqnarray}
The QED contribution is known up to the four-loop order (plus the dominant
terms from five loops) and $a_\mu^{\rm ew}$ is computed up to the
two-loop level:
\begin{eqnarray}
a_\mu^{\rm QED} \,\,=\,\, 116\,584\,705.6\pm2.9\times10^{-11}\,,
&\quad&
a_\mu^{\rm ew} \,\,=\,\, 151\pm4\times10^{-11}
\,.
\end{eqnarray}
Both contributions exhibit a small error and are thus completely under
control. The hadronic contribution again has to be obtained via a dispersion
relation
\begin{eqnarray}
a_\mu^{\rm had} &=& \frac{\alpha^2(0)}{3\pi^2}
\int_{4m_\pi^2}^\infty\,{\rm d}
s\,\frac{K(s)}{s} R(s)
\,,
\label{eqamu}
\end{eqnarray}
where $K(s)$ denotes the QED kernel function.
$K(s)$ is strongly peaked for small values of $\sqrt{s}$. Actually 98\% of the
contribution arises from energies $\sqrt{s}<1.8$~GeV.

Let us now discuss the various improvements which intend to reduce the
error on $\Delta\alpha_{\rm had}^{(5)}$ and $a_\mu^{\rm had}$.

In Ref.~\cite{AleDavHoe97} $\tau$ data from ALEPH has been used in order to
get more information about $R(s)$ for energies below roughly 1.8~GeV.
The hypothesis of conserved vector current (CVC) in combination with isospin
invariance relates, e.g.,  the vector part of the two-pion $\tau$ spectral
function to the corresponding part of the isovector $e^+e^-$ cross section
through the following relation
\begin{eqnarray}
\sigma^{I=1}\left(e^+e^-\to\pi^+\pi^-\right) &=&
\frac{4\pi\alpha^2(0)}{s}v_{J=1}\left(\tau\to\pi\pi^0\nu_\tau\right)
\,.
\end{eqnarray}
A similar equation holds for the four-pion final state.
Their incorporation into the analysis has been performed in~\cite{AleDavHoe97}
leading to a slight reduction of the error on $\Delta\alpha^{(5)}_{\rm had}$,
however, to a significant reduction for $a_\mu^{\rm had}$ (see table below).
In the meantime new date from the Novosibirsk experiment CMD-2 became
available~\cite{JTho}. The main improvement results from
the reduction of the
systematic error from 2 to 1.5\% in the energy region below 1.4~GeV.
This brings the error for $a_\mu^{\rm had}$ (obtained from $e^+e^-$ data
only) down
to the same order of magnitude as the one
cited in~\cite{AleDavHoe97}.
However, it remains to check if also the central values are in agreement.

Looking at the data in the energy region above $2$~GeV one realizes that they
are accompanied with large systematic uncertainties. Thus there is the
temptation to replace the inaccurate data by predictions from 
perturbative QCD (pQCD)
at least in those energy regions which are not affected by
resonance phenomena.
This is also supported from recent QCD analyses performed by
ALEPH~\cite{ALEPHtau} and OPAL~\cite{OPALtau} using hadronic $\tau$
decays. Not only for the
total rate but also for the spectral function towards the upper 
end the substitution of imprecise data by pQCD seems
justified. 

$R(s)$ can be calculated in the framework of
pQCD up to order $\alpha_s^3$ if quark masses are neglected and up to
${\cal O}(\alpha_s^2)$ with full quark mass dependence
(see Ref.~\cite{CheHoaKueSteTeu97} and references therein).
In Ref.~\cite{DavHoe98_1} pQCD has been used down to an energy scale of
$\sqrt{s}=1.8$~GeV and it has been shown that the non pertubative
contributions are small. This leads to a further reduction of the error of
about a factor two.

In~\cite{KueSte98} pQCD also has been applied down to small center-of-mass
energies implementing the state-of-the-art corrections up to order
$\alpha_s^3$. The full charm and bottom quark mass effects are taken into
account at two-loop order. 
All formulae are available for arbitrary renormalization scale $\mu$
which allows to test the scale dependence of the final answer. This
has been used to estimate the theoretical uncertainties
from uncalculated higher orders.
The details of the formalism can be found in~\cite{CheHoaKueSteTeu97}.

Perturbative QCD is clearly inapplicable in the charm
threshold region
between $3.7$ and $5$~GeV
where rapid variations of the cross section are
observed. Data have been taken more than $15$ years ago by the 
PLUTO, DASP, and MARK~I collaborations.
The systematic errors of $10$ to
$20$\% exceed the statistical ones significantly.
They are reflected in a
sizeable spread of the experimental results.
In~\cite{KueSte98} the experimental data are normalized to match the
predictions of perturbative QCD both below $3.7$ and above $5.0$~GeV.
Two models have been constructed which describe the differences of the
normalization factors below and above the considered energy interval.
Similar statements hold for the bottom threshold region in the range from
$10.5$~GeV to $11.2$~GeV. There, however, the numerical contribution is much
less significant.

A different approach for the evaluation of $\Delta\alpha_{\rm had}^{(5)}$,
based on QCD sum rules (SR),
has been used in~\cite{GroKoeSchNas98}.
Global parton-hadron duality is used in order to reduce the influence of
the data in the different intervals.
This is achieved by choosing a proper polynomial, $Q_N(s)$,
which is supposed to 
approximate the weight function $M_Z^2/s(M_Z^2-s)$ as good as possible.
Adding and subtracting $Q_N(s)$ in Eq.~(\ref{eq:delaldisp})
and exploiting the analyticity of the
subtracted term leads to
\begin{eqnarray}
\int_{s_0}^{s_1}\,{\rm d}s\,
\frac{R(s)}{s\left(s-M_Z^2-i\epsilon\right)}
&=&
\int_{s_0}^{s_1}\,{\rm d}s\,
R(s)\left(\frac{1}{s\left(s-M_Z^2-i\epsilon\right)}-Q_N(s)\right)
\nonumber\\&&\mbox{}\qquad\qquad
+
6\pi i\oint_{|s|=s_1} \!\!\!{\rm d}s\, \Pi^{\rm QCD}(s) Q_N(s)
\,.
\label{eq:delalsub}
\end{eqnarray}
Thus the influence of the experimental data is significantly reduced in the
first term of the r.h.s. and pQCD only has to be used for 
$|s|=s_1$ which is indicated by the superscript ``QCD''.
In~\cite{GroKoeSchNas98} the interval $[2m_\pi,40~\mbox{GeV}]$ has been
subdivided into four parts leading to the values $\sqrt{s_1}=3.1$~GeV,
$9.46$~GeV, $30$~GeV and $40$~GeV, respectively, for the upper integration
bound.
This subdivision is necessary as 
for large $N$ more and more stress is put on the
unknown QCD-input in the second term of Eq.~(\ref{eq:delalsub}).
The authors of~\cite{DavHoe98_1} applied similar methods to the energy 
intervals $[2m_\pi,1.8~\mbox{GeV}]$ and $[3.7~\mbox{GeV},5.0~\mbox{GeV}]$
in order to improve their previous analysis~\cite{DavHoe98_2}.

An approach complementary to the ones mentioned above has been chosen
in~\cite{Erl98}. It is based on unsubtracted dispersion relations (UDR)
which are used in order to evaluate the electromagnetic coupling in the
$\overline{\rm MS}$ scheme. For the energy region below $1.8$~GeV
the data analysis of~\cite{DavHoe98_1} is adopted. Then four-loop running is
accompanied by three-loop matching in order to arrive at $\bar\alpha(M_Z^2)$,
which subsequently has to be transformed to the on-shell quantity
$\alpha(M_Z^2)$. Via this method no complications in connection with the
$J/\Psi$ or $\Upsilon$ resonances occur. However, one encounters a much
stronger dependence on the quark masses.
\begin{table}[t]
\renewcommand{\arraystretch}{1.}
\begin{center}
\caption{\label{tab:cmp}
Comparison of the recent improvements on the error of
$\Delta\alpha^{(5)}_{\rm had}(M_Z^2)$ and $a_\mu^{\rm had}$ with the
values given in~\protect\cite{EidJeg95}.
The column ``comment'' reminds on the different methods used in the
analysis as described in the text.
(${}^*\Delta\alpha_{\rm top}(M_Z^2)$ subtracted;
${}^{**}$value corresponding to $\alpha_s(M_Z^2)=0.118$ adopted.)
}
{\small
\begin{tabular}{|l|l|l|l|}
\hline
$\Delta\alpha^{(5)}_{\rm had}(M_Z^2)\times 10^4$ 
& $a_\mu^{\rm had}\times10^{-11}$
& Ref.
& comment
\\
\hline
$280 \pm 7$
& $7024\pm153$
& \cite{EidJeg95}
& data\\
\hline
$281.7 \pm 6.2$   
& $7011\pm94$
& \cite{AleDavHoe97}
& $\tau$ data\\
$278.4 \pm 2.6^*$   
& $6951\pm75$
& \cite{DavHoe98_1}
& + pQCD\\
$277.5 \pm 1.7$   
& ---
& \cite{KueSte98}
& + ``charm threshold''\\
$277.6 \pm 4.1$   
& ---
& \cite{GroKoeSchNas98}
& SR\\
$277.3 \pm 2.0^{**}$   
& ---
& \cite{Erl98}
& $\tau$ data + UDR\\
$277.0 \pm 1.6^*$   
& $6924\pm62$
& \cite{DavHoe98_2}
& $\tau$ data + pQCD + SR\\
\hline
\end{tabular}
}
\end{center}
\end{table}

In Tab.~\ref{tab:cmp} the recent evaluations of $\Delta\alpha^{(5)}_{\rm
  had}(M_Z^2)$ and $a_\mu^{\rm had}$ are compared with the one of
Ref.~\cite{EidJeg95}. A significant reduction of the error is observed.
It is very impressive that the new analysis show very good agreement both in
their central values and in their quoted errors.
This development is very promising and the new values should be considered in
the analysises of the precision data. Once more precise experimental input is
available it can replace the theory-motivated parts in
Refs.~\cite{AleDavHoe97,DavHoe98_1,KueSte98,GroKoeSchNas98,Erl98,DavHoe98_2}.

\section*{Acknowledgments}
I would like to thank J.H.~K\"uhn for the fruitful collaboration on this
subject and the organizers of WIN 99 for the very nice atmosphere during the
workshop. This work was supported by the {\it Schweizer Nationalfond}.

\section*{References}


\end{document}